\documentclass[prb,twocolumn,floatfix,showpacs,amsmath,amssymb]{revtex4}
\usepackage{bbm}
\usepackage{amsfonts}
\usepackage{graphicx}
\usepackage{color}
\usepackage{amsmath}
\usepackage{amssymb}
\usepackage{latexsym}
\usepackage{psfrag}

\begin{document}

\title{Interactions and screening in gated bilayer graphene nanoribbons}
\author{Hengyi Xu}
\email{hengyi.xu@uni-duesseldorf.de}
\author{T. Heinzel}
\affiliation{Condensed Matter Physics Laboratory, Heinrich-Heine-Universit\"at,
Universit\"atsstr.1, 40225 D\"usseldorf, Germany}
\author{A. A. Shylau}
\author{I. V. Zozoulenko}
\email{igor.zozoulenko@itn.liu.se}
\affiliation{Solid State Electronics, Department of Science and Technology, Link\"{o}ping
University, 60174 Norrk\"{o}ping, Sweden}
\date{\today }

\begin{abstract}
The effects of Coulomb interactions on the electronic
properties of bilayer graphene nanoribbons (BGNs) covered by a gate
electrode are studied theoretically. The electron density distribution and
the potential profile are calculated self-consistently within the Hartree
approximation. A comparison to their single-particle counterparts reveals
the effects of  interactions and screening. Due to the finite width of
the nanoribbon in combination with electronic repulsion, the gate-induced
electrons tend to accumulate along the BGN edges where the potential assumes
a sharp triangular shape. This has a profound effect on the energy gap
between electron and hole bands, which depends nonmonotonously on the gate
voltage and collapses at intermediate electric fields. We interpret this behavior in terms of interaction-induced warping of the energy
dispersion.
\end{abstract}

\maketitle

\section{Introduction}

Graphene and its bilayers are extremely good conductors with mobilities up
to $20\mathrm{m^2/Vs}$ at room temperature, \cite{Morozov2008,Chen2008}
which makes them interesting as active channels in field-effect transistors
(FETs). These materials are also considered as building blocks for quantum
electronics such as quantum computation and pseudospintronic devices. \cite%
{Katsnelson2006,Trauzettel,San-Jose2009} To facilitate these applications,
large on-off resistance ratios are necessary. \cite{Pereira2009} Several
mechanisms for opening up an energy gap in two-dimensional (2D) monolayer
graphene have been proposed to achieve this goal. \cite{Filho2007} It has
been suggested that the application of an electric field between the two
graphene layers breaks the inversion symmetry and induces an energy gap
between electron and hole bands. \cite{McCannprl2006} Oostinga et al. \cite%
{Oostinga} have recently been able to detect such an electric-field induced
energy gap in a graphene bilayer system. Infrared absorption spectroscopy
furthermore demonstrated a gate-controlled band gap with gap energies up to $%
250\,\mathrm{meV}$, which is in accordance with the self-consistent
tight-binding calculations. \cite{Zhang2009,Mak2009} The size of the gap is
determined by the electric field between the two layers which can be tuned
in single- \cite{Xia2010} or double- \cite{Koshino2009a} gate geometries.
Recently, the influence of a potential applied to a single gate on bilayer
graphene has been studied theoretically within the continuum model,
revealing a roughly linear dependence of the gap width as the carrier
density increases. \cite{McCann2006} However, a nonmonotonous evolution of
the gap width with the electron density has been predicted for graphene triple
and quadruple layer systems, which has been attributed to trigonal warping
of the band structure. \cite{Avetisyan2009_1,Avetisyan2009_2} Moreover,
ab-initio density functional theory calculations have been used extensively
to investigate the electronic structure of bilayer graphene and essentially
confirm the behavior suggested by the tight-binding and continuum models.
\cite{Min2007,Yu2008,Gava2009}

Due to the presence of edges, graphene nanoribbons (GNRs) reveal a much
richer phenomenology than 2D graphene sheets. \cite%
{Han2007,Wang2008,Areshkin2007} GNRs possess energy gaps due to size
quantization and edge states are formed with properties depending on the
width and the type of edges. \cite{Wakabayashi1999,Nakada1996} Edge disorder
may result in transport gaps around the neutrality point. \cite%
{Han2007,Evaldsson2008,Xu2009,Mucciolo2009} For gated GNRs, the electronic
structure of monolayer systems has been modeled by Fernandez-Rossier with
the image charge method in the Hartree approximation. \cite%
{Fernandez-Rossier2007} The inhomogeneous charge density and potential
across the nanoribbons were found. Also, the small transverse size has a strong
impact on the classical and quantum capacitances of monolayer GNRs. \cite{Shylau2009}

It is thus self-evident that interactions and  imperfect
screening should influence the electronic properties in bilayer GNRs (which we will refer to as BGNs below) as well and may modify the formation of the gate-voltage induced energy gap. However, neither the gate electrostatics, nor the self-consistent band structure and the band gap formation in  BGNs (in contrast to bulk bilayers) have been studied before. In the present paper, we address this issue by taking Coulomb interactions in
BGNs  into account and studying the electron density distribution, the Hartree potential and the energy dispersion, focusing
particularly on the evolution of the electric field induced band gap which can be
regarded as a single-valued parameter for the relevance of the interaction effects.

The paper is organized as follows. In Sec. II we introduce the structure to
be investigated and formulate our model. The results are presented and
discussed in Sec. III. Sec. IV contains a summary and conclusions.

\section{Structure and Model}

Throughout this work we consider long BGNs stacked in the Bernal form with
ideal armchair edges. We note that the type of edge of the BGN has only a
marginal influence on our results and that finite-lengths effects are
neglected. The tight-binding Hamiltonian reads \cite{Neto2009}
\begin{widetext}
\begin{eqnarray*}
H &=&\sum_{\ell ,\langle i,j\rangle }(V_{\ell ,i}a_{\ell ,i}^{+}a_{\ell
,i}+V_{\ell ,j}b_{\ell ,j}^{+}b_{\ell ,j})-\gamma _{0}\sum_{\ell ,\langle
i,j\rangle }(a_{\ell ,i}^{+}b_{\ell ,j}+h.c.)
\end{eqnarray*}
\begin{eqnarray}  \label{eqhamiton}
 -\gamma _{1}\sum_{i}(a_{1,i}^{+}b_{2,i}+h.c.)-\gamma _{3}\sum_{\langle
i,j\rangle }(b_{1,i}^{+}a_{2,j}+h.c.)
\end{eqnarray}
\end{widetext}
where $a_{\ell ,i}^{+}$ ($b_{\ell ,i}^{+}$) are the creation operators at
sublattice $A$ ($B$) in layer $\ell =1,2$ at site $\mathbf{r}_{i}$. We adopt
the common graphite nomenclature and include the three coupling energies $%
\gamma_0$, $\gamma_1$ and $\gamma_3$, where $\gamma _{0}=3.16\, \mathrm{eV}$
represents the intra-layer nearest-neighbor coupling energy, $%
\gamma_{1}=0.39\, \mathrm{eV}$ is the coupling energy between sublattice $B$
and $A^{\prime}$ in different graphene layers, and $\gamma_{3}=0.315\, \mathrm{eV}$ the hopping energy between sublattice $A$ and $B^{\prime}$ in the lower
and upper layers, respectively. $V_{\ell}$ is the the on-site potential
arising from the Coulomb interaction of the induced charges the density of
which is to be calculated.

\begin{figure}[tbp]
\includegraphics[scale=1]{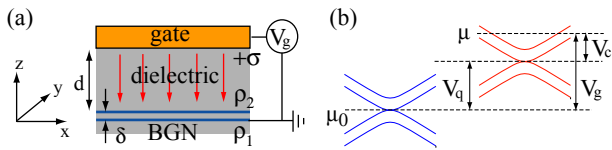} 
\caption{(a) Scheme of the BGN device. The arrows indicate the direction of
electric field for $V_g>0$.(b) Illustration of the band structure shift
induced by the gate voltage $V_g$, and of the voltage components $V_c$ and $%
V_q$. }
\label{fgfet}
\end{figure}

In the capacitor structure shown in Fig. \ref{fgfet}(a), a BGN of width $w$
in y-direction and layer spacing $\delta=0.335\,\mathrm{nm}$ in z-direction extends
along the x-direction and is embedded in an insulating medium (we assume a
relative permittivity $\varepsilon_r$=3.9 adequate for SiO$_2$). The
structure is covered by a homogeneous metallic gate extending in the xy -
plane at distance $d$. Electrical access is enabled by connecting a voltage source to the metal gate and the lower BGN layer. A neutral BGN is obtained by setting
the on-site potential to zero, corresponding to the Fermi energy $E_F=0$. As
a gate voltage $V_g$ is applied, excess charges $\rho _{i}=en_{i}$  are induced in the lower (i=1) and upper (i=2) graphene layer summing up to the total charge density $\rho=en$. Here, $n_{i}$ denotes the carrier density in layer i and $e$ the electron charge. This  is balanced by the charge density $\sigma $ of opposite polarity at the metal-oxide interface such that the
Fermi energy departs from the neutrality point, while the charge neutrality
of the whole system is conserved. In the following calculations, we subtract
$V_g$ from both the gate electrode as well as from the BGN which does not
modify the electrostatics of the problem. This way, the gate metal-insulator
interface is kept grounded which allows us to take advantage of the image
charge technique creating a charge density at a distance $d$
above the metal-insulator interface. \cite{Peres2009}

Within our framework, $eV_{g}$ is defined as the difference between the
electrochemical potential $\mu $ and the neutrality point of the discharged BGN $%
\mu _{0}$,
\begin{equation}
eV_{g}=\mu -\mu _{0},  \label{eqchep}
\end{equation}%
see Fig. \ref{fgfet} (b) for an illustration. The displacement of the chemical
potential is composed of the classical term $eV_{c}$ associated with the
electrostatic potential, and of the quantum term $eV_{q}$ describing the
accumulation of extra charges in the energy bands,
\begin{equation}
V_{g}=V_{c}+V_{q}.  \label{eqvg}
\end{equation}%

The charge distribution screens the lateral electric field components at the
gate. Taking the Coulomb interactions between the charges and their images
into account, the Hartree potential at the site $\mathbf{r}$ in the
continuous limit can be computed as
\begin{widetext}
\begin{equation}  \label{eqhartv}
V(\mathbf{r}(x,y))=-\frac{e}{4\pi \varepsilon_0\varepsilon_r}\int dy^{\prime
}\rho(y^{\prime})\left( \ln\frac{(y-y^{\prime})^2}{(y-y^{\prime})^2+4(d+\xi
\delta)^2}+\ln\frac{(y-y^{\prime})^2+\delta^2}{(y-y^{\prime})^2+(2d+\delta)^2}%
\right)
\end{equation}
\end{widetext}
with $\xi=1$ and $0$ for the lower and the upper layer of the BGN, respectively. The first term describes the interaction between the
charges in each layer and their corresponding mirror charges, while the
second term accounts for the interaction of the charges in one layer with
the mirror charges related to the other layer.

In order to determine the charge density, the local density of states $D(r,E)$ (LDOS) at site $r$ and energy $E$ is
calculated with the help of the real-space Green's functions technique
described in Ref. \onlinecite{Xu2008}. The electron density at site $r$ is then given by
\begin{equation}
n (r)=\int_{-\infty }^{\mu }D_{sc}(E)f(E)dE-\int_{-%
\infty }^{\mu _{0}}D_{0}(E)f(E)dE  \label{eqrho}
\end{equation}
where $f(E)$ is the Fermi-Dirac distribution. $D_{sc}(E)$ and $
D_{0}(E)$ represent the LDOS at energy $E$ with and without
including the self-consistent interactions, respectively. In the actual
calculations, the second term is replaced by $4/\sqrt{3}a^{2}$ with $%
a=0.246\,\mathrm{nm}$ which represents the positive charge background of the ions. Since
the Hartree potential, Eq.  (\ref{eqhartv}) depends on $n (r)$
which is a solution of the Schr\"{o}dinger equation with the Hamiltonian given by Eq.
(\ref{eqhamiton}), these equations are solved by the Broyden iterative method.

Some insight into the basic physics of gated BGNs is obtained from the
analytical expression for the potential difference between the layers
induced by gate voltage. Since $\delta \ll d$, we can approximate the BGN by
a conducting strip with the total charge density $\rho$. The relation between gate voltage and the charge density for the
system formed between the conducting strip of width $w$ located at the
distance $d$ apart from a semi-infinite plate is given by \cite{Shylau2009}
\begin{equation}
V_{g}=\frac{\rho }{\pi \varepsilon _{0}\varepsilon _{r}}\left[ 2d\arctan
\left( \frac{w}{4d}\right) +\frac{w}{4}\ln \left\{ 1+\left( \frac{4d}{w}%
\right) ^{2}\right\} \right]  \label{vgn}
\end{equation}%
The potential difference between the graphene layers can be calculated by
integrating the electric field in between, i.e., $\Delta
V=\int_{d}^{d+\delta }E(z)dz$. By using the method of image charges, one
finds that $E(z)$ is determined by
\begin{widetext}
\begin{equation}
E(z)=E_{\rho _{2}}^{strip}(z+d)+E_{\rho
_{1}}^{strip}(z+(d+\delta ))+E_{-\rho _{2}}^{strip}(z-d)+E_{-\rho
_{1}}^{strip}(z-(d+\delta ))
\end{equation}
\end{widetext}
where the first two terms in the summation
correspond to the image charges. The intensity of the electric field at
distance $z$ from the middle of a strip is given
by \cite{Shylau2009}
\begin{equation}
E_{\rho }^{strip}(z)=\frac{\rho }{\pi \varepsilon _{0}\varepsilon _{r}}%
\arctan \left( \frac{w}{2z}\right)  \label{Estrip}
\end{equation}%
Performing the integration and using Eqs. (\ref{vgn})-(\ref{Estrip}), we get
\begin{equation}
\Delta V=\frac{\delta \arctan \left( \frac{w}{4d}\right) V_{g}}{\left[
2d\arctan \left( \frac{w}{4d}\right) +\frac{w}{4}\ln \left\{ 1+\left( \frac{%
4d}{w}\right) ^{2}\right\} \right] }-\frac{\Delta \rho }{2\varepsilon
_{0}\varepsilon _{r}}  \label{eqdv}
\end{equation}%
where $\Delta \rho =\rho _{2}-\rho _{1}$ is the charge density difference.
In the derivation of Eq. (\ref{eqdv}), a homogeneous distribution of the
charge density in the BGN has been assumed. This assumption is a reasonable
approximation for low charge densities.

\section{Results and Discussion}

\begin{figure}[tbp]
\includegraphics[scale=1]{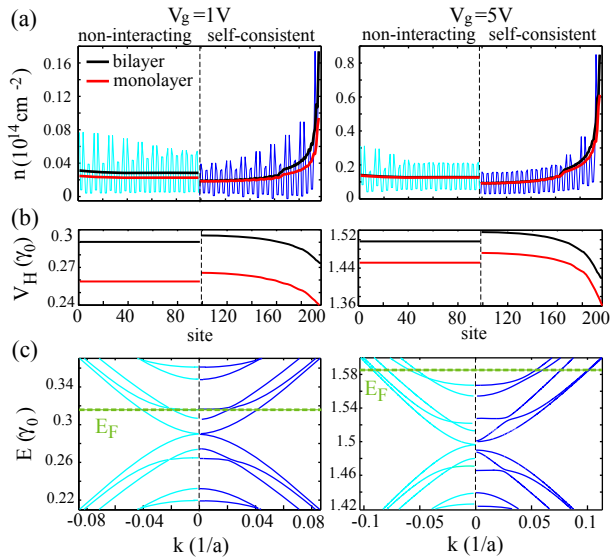} 
\caption{Comparison between the single-particle (left parts) and the
self-consistent Hartree (right parts) calculations for different gate
voltages. The distance between the gate and the BGN is set to $d=100\,\mathrm{nm}$. (a) Induced electron density distribution along
the $y$ direction (thin lines) and  the values averaged over six neighbors (bold lines).
(b) Potential profiles from the one-electron and the
Hartree approximation calculations,  averaged over six neighbors. (c) Band structures. The dashed horizontal lines indicate the Fermi level.}
\label{fgrhopot}
\end{figure}

Figs. \ref{fgrhopot}(a) and (b) show the electron density distributions and
Hartree potentials of BGN with $N=49$ sites in transverse direction,
corresponding to $w=12\mathrm{nm}$ for different gate voltages. The
calculations are performed over one unit cell containing $4N$ sites. To
clarify the role of electron-electron interactions we show the
self-consistent Hartree (right parts) and non-interacting (left parts)
calculations. In our context, non-interacting calculations correspond to the
Hamiltonian (\ref{eqhamiton}) where the potential difference between the
layers is given by Eq. (\ref{eqdv}) with  $\Delta \rho $ set to
0.  Both non-interacting
and self-consistent calculations show pronounced oscillations between
neighboring sites. Because of this we also plot for clarity the density
distributions averaged over six neighboring sites. Within the
self-consistent treatment, the electrons tend to accumulate along the edges
as a manifestation of electronic repulsion. As the gate voltage is increased, the edge
accumulation becomes more prominent. It is worth pointing out
that it is primarily the applied gate voltage which determines this
distribution, resembling that one found in monolayer GNRs. \cite{Silvestrov2008,Shylau2009} This edge accumulation
is reflected in the Hartree potential as the formation of triangular-shaped
wells near the edges, an effect known in quantum wires prepared by
cleaved-edge overgrowth \cite{Ihnatsenka2006} as well as from wide
two-dimensional quantum wells \cite{Suen1991}. In the non-interacting case,
on the other hand, the charge density distributions which are obtained by
using the potential profiles in the left panels are quite flat, while the
electron density oscillations between neighboring sites persist.

For comparison, we also plot the electron density and potential of a
monolayer GNR of identical width. The constant potentials shown in the left
panels are classical responses of the BGN to the gate voltage corresponding
to the position of the charge neutrality point $V_c$. The BGN system can be
viewed as two monolayer GNRs connected in series such that the classical
capacitance of the BGN is smaller than that of monolayer GNR, resulting in
larger values of $V_c$. Comparing the averaged electron density, one
finds that they are very close at the center of nanoribbons in both cases,
which is consistent with the fact that the total induced charge density
equals to the sum of the densities in each layer, i.e., satisfying the
relation $\sigma=-(\rho_1+\rho_2)$ according to electrostatics for bilayer.
There are however some small differences close to edges which we interpret
as a consequence of interlayer Coulomb interactions.

Fig. \ref{fgrhopot}(c) shows the dispersion relations of $V_g=1 V$ and $5 V$
for the non-interacting and self-consistent cases. Within the
single-particle picture, the potential energy across the BGN is constant and
the dispersion relations are shifted upwards almost rigidly as the gate
voltages increases. The energy gaps induced by the gate voltage are barely
visible. Turning on the electronic interactions, the potential profile
changes only slightly for $V_g=1V$. However, as the gate voltage increases
to $5V$, the dispersion relation gets strongly modified. Anticrossings are
generated and a significant energy gap is opened between the conduction and
valence bands, originating not only from the potential difference between
the layers but also from the transverse potential profile. We note in this respect that even in
monolayer GNRs, the external electric field can lead to a gap modulation for
semiconductor armchair nanoribbons \cite{Ritter2008} and to a gap opening
for zigzag ribbons \cite{Son2006}.

\begin{figure}[tbp]
\includegraphics[scale=1]{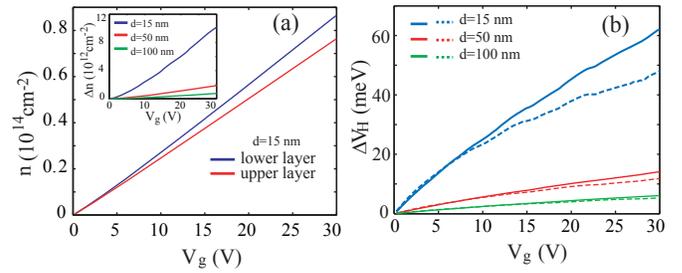} 
\caption{(a) The average electron density on each graphene layer for
dielectric thickness $d=15 \mathrm{nm}$. Inset: The difference of the
average density between layers for different dielectric thicknesses. (b) The
difference of Hartree potentials between graphene layers. Solid and dashed
lines correspond to the numerical and analytical calculations from Eq. (%
\protect\ref{eqdv}), respectively.}
\label{fgdrho}
\end{figure}

We continue by considering the average electron density and the Hartree
potential in each graphene layer as a function of the gate oxide thickness $%
d $. As shown in Fig. \ref{fgdrho}(a), the electron densities in both layers
increase approximately linearly $V_g$. In monolayer GNRs, a similar linear relation of electron
density and gate voltage $\rho\propto V_g$ has been reported. \cite%
{Fernandez-Rossier2007} This common feature indicates that the classical
electrostatic contribution predominantly determines the charge density.
Furthermore, the electron densities of the two layers of the BGN diverge as
the gate voltage increases. This electron density difference grows monotonously as shown in the inset. The dependence of the
potential differences on $d$ is basically governed
by the classical component. Fig. \ref{fgdrho}(b) shows the corresponding
results of the Hartree potential. The analytical expression according to Eq. (\ref{eqdv})
quantitatively reproduces the exact numerical results at the low gate
voltages, especially for larger dielectric thickness $d=50\,\mathrm{nm}$ and $%
100\,\mathrm{nm}$. However, deviations emerge for higher gate voltages, most
pronounced for $d=15\,\mathrm{nm}$. They can be understood via the derivation
of Eq. (\ref{eqdv}), which is based on the assumption that induced electron
density is homogeneous and the potential across the ribbon is constant. In
reality, both a larger $V_g$ and a smaller $d$ lead
to the higher electron densities, stronger Coulomb repulsion and more
pronounced potential inhomogeneities, rendering the classical capacitor
approximation less valid.

\begin{figure}[tbp]
\includegraphics[scale=1]{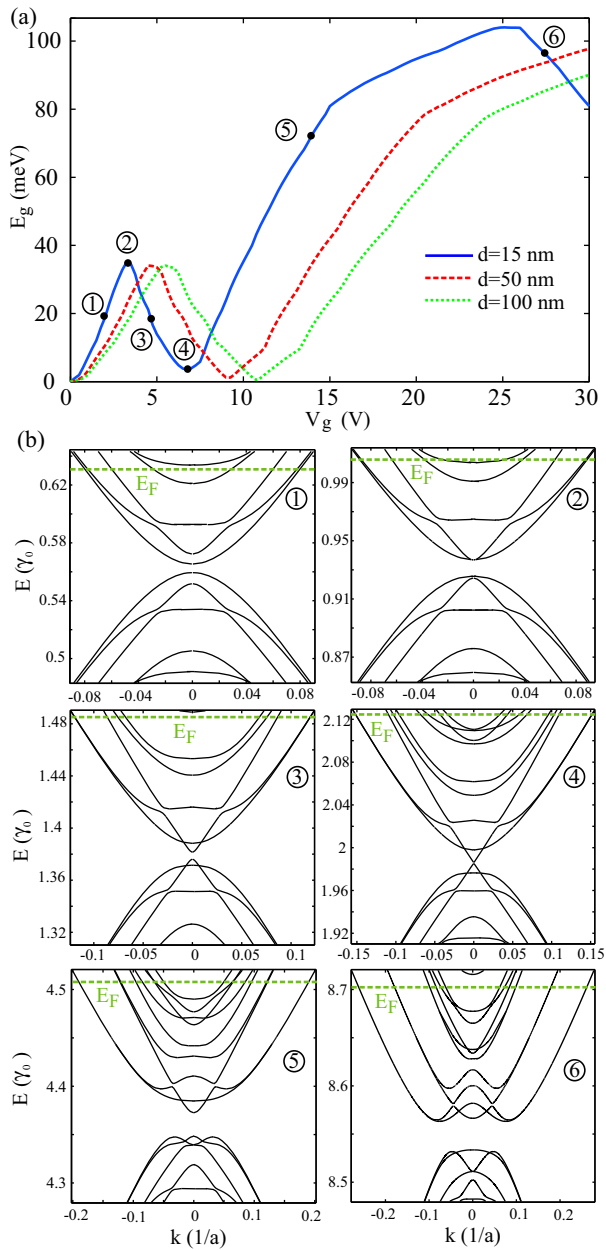} 
\caption{(a) The energy gap as a function of the applied gate voltage for
various dielectric thicknesses $d$. (b) Representative band structures
corresponding to the Fermi energies shown in (a). The dashed lines indicate
the positions of Fermi energy. }
\label{fgeg}
\end{figure}

We proceed with a detailed study of the effects of Coulomb interactions on the band
structures, and begin with looking at the size of the energy gap as a function of the
gate voltage for different dielectric thicknesses as shown in Fig. \ref{fgeg}(a). In
contrast to the 2D bilayer graphene, where the band gap increases monotonously as the
applied electric field is increased, \cite{Zhang2009} the energy gap in BGNs exhibits a
pronounced dip at low gate voltages for all values of $d$. If the single-particle
potential is used instead, the energy gap increases monotonously with $V_{g}$ and follows
closely the value of the Hartree potential difference shown in Fig. \ref{fgdrho}(b), and
the dip is absent. This implies that the interactions are responsible for the formation
of the dips. To shed light on the origin of this suppression of the band gap formation,
let us inspect the band structures for $d=15\mathrm{nm}$ at
several representative Fermi energies. For the first two selected values of $%
V_{g}$, \textcircled{1} and \textcircled{2}, the size of gap is proportional
to $V_{g}$. One observes that the second energy band of the electrons
develops an anticrossing with the third one and develops a strong curvature
around $k_{x}=0$. The minimum of this band approaches the conduction band
edge formed by the lowest energy band and takes over the role of the band
edge at the scenarios depicted in Figs. \ref{fgeg}(b), \textcircled{3} and %
\textcircled{4}. As a result, the energy gap gets reduced as the gate
voltage increases and can even approach zero for some parameters. We note
that a similar effect has been discussed by Avetisyan et al., in graphene
tri- and quadruple-layer systems based on a different mechanism, however.
\cite{Avetisyan2009_1} As the voltage is increased beyond the dip of
the energy gap (e.g. $V_{g}\gtrsim 25$ for $d=15$  nm),
the familiar Mexican hat shaped dispersion emerges which leads to the
reduction of the energy gap while the electron-hole symmetry vanishes.

From this evolution of the energy band dispersions, it becomes apparent that
the Hartree term induces some warping of the energy dispersion. We note that in contrast to 2D
bilayer systems, \cite{Koshino2009} this effect is not due to $\gamma_3$,
since after setting $\gamma_3 =0$ the band structure changes only slightly
but the collapse of the band gap persists (not shown here). The prerequisite for is the
potential inhomogeneity generated by the charge accumulation at the edges. This is why the collapse of the band gap is absent in the 2D cases. Furthermore, the small reduction
of the energy gap for large $V_g$ in the system with $d=15 \,\mathrm{nm}$ is
not directly related to warping but is the result of the Mexican-hat band
formation.

\section{Summary and Conclusions}

We have studied the electron density distribution and Hartree potential of
a single-gated BGN. Coulomb interactions are incorporated in a self-consistent way within the Hartree approximation. The repelling electrons accumulate at the BGN edges and induce characteristic dips in the transverse potential profile. They change the band structures strongly and modify the energy gaps which thereby behave qualitatively differently compared to 2D bilayer systems, showing an unusual reduction in the sizes of the energy gaps at intermediate gate voltages. An analytical expression for the potential difference between the two layers is obtained based on the assumption of a classical capacitor and compared with
self-consistent numerical calculations. They show good agreement at small
gate voltages. The discrepancies found at higher voltages and small dielectric
thicknesses are due to the assumption of homogeneous charge density and
constant potential in the analytical model. This becomes invalid for high
electron densities because the quantum mechanical effects modify
the charge redistribution significantly.\\
A disadvantage of the single-gate structure discussed here is that due to
small capacitance, the Fermi energy is far above the neutrality point in the regime of the collapse of the band gap, such that this effect is not directly accessible experimentally.  This shortcoming can conceptually be overcome by performing corresponding simulations for a double-gate structure, where independent tuning of the Fermi energy and the interlayer electric field is possible.\cite{Oostinga,Koshino2009a}

\begin{acknowledgments}
H.X. and T.H. acknowledge financial support from Heinrich-Heine-Universit\"{a}t D\"{u}sseldorf and the German Academic Exchange Service (DAAD) within
the DAAD-STINT collaborative grant. A.A.S. and I.V.Z. acknowledge the
support from the Swedish Research Council (VR) and from the Swedish
Foundation for International Cooperation in Research and Higher Education
(STINT) within the DAAD-STINT collaborative grant.
\end{acknowledgments}

\end{document}